\pdfoutput=1
\documentclass[traditabstract,rnote]{aa}
\usepackage{graphicx}
\usepackage{txfonts}
\begin{document}
\title{On Broyden's method for the solution of the multilevel non--LTE
  radiation transfer problem}

   \author{S. Nicolas
          \and
          L. Bigarr\'e
          \and
          F. Paletou
          }

          \institute{Institut de Recherche en Astrophysique et Plan\'etologie, Universit\'e de Toulouse, CNRS, 14 av. E. Belin, 31400 Toulouse, France\\
            \email{fpaletou@ast.obs-mip.fr} }

   \date{Received October 13, 2010; accepted December 12, 2010}


   \abstract {This study concerns the fast and accurate solution of multilevel non-LTE radiation transfer problems. We propose and evaluate an alternative iterative scheme to the classical MALI method. Our study is indeed based on the application of Broyden's method for the solution of nonlinear systems of equations. Comparative tests, in 1D plane-parallel geometry, between the popular MALI method and our alternative method are discussed. The Broyden method is typically 4.5 times faster than MALI. It makes it also fairly competitive with Gauss-Seidel and Successive Over-Relaxation methods developed after MALI.}


   \maketitle

%

\section{Introduction}

The solution of the non-LTE multilevel-atom radiative transfer problem
is a classical one in astrophysics. Indeed, the assumption of non-LTE
implies, consistently with the departure of the source functions from
Planck functions, that the population density of the atomic or
molecular levels considered depart from what can be derived at LTE, in
a straightforward manner, using Saha and Boltzmann relations (see
e.g., Mihalas 1978).

In the non-LTE case, one has on the contrary to solve simultaneously
and self-consistently for a set of $N_{\rm T}$ equations of radiative
transfer together with $N_{\rm L}$ equations of statistical
equilibrium (hereafter ESE) describing the detailed balanced between
excitation and de-excitation processes between every atomic or
molecular levels. Since absorption and stimulated emission radiative
rates depends explicitely on the radiation field, which itself depends
on the level populations, this problem is intrinsically a search for
the solution of coupled {\em nonlinear} equations.

Since the beginning of numerical radiative transfer in the late 60's,
the two most popular methods used for tackling this problem have been
the complete linearization method of Auer \& Mihalas (1969) and the
Accelerated $\Lambda$-Iteration based scheme called MALI (Rybicki \&
Hummer 1991). Despite their apparent differences, they have however in
common the fact that basically, one is conducted to deal with {\em
  linearized} equations. An interesting comparative study of these two
approaches have been made by Socas-Navarro \& Trujillo Bueno
(1997).

In this study, we investigate on the use of a quasi-Newton numerical
method for the solution of the nonlinear ESE. Our choice went to
Broyden's method (1965) whose elements will be presented in \S2.  To
the best of our knowledge, Koesterke et al. (1992) were the first to
bring this numerical scheme into the field of radiation
transfer. Their study was presented in the context of the modelling of
spherically expanding atmospheres of hot and massive Wolf-Rayet
stars. Broyden's method was more recently invoked in the context of
the coupled-escape probability method (Elitzur \& Asensio Ramos 2006).

Besides from the required algebra and mention to caveats related to
the implementation of the method, it remains however difficult to
figure out from Koesterke et al. (1992) the actual performances of
such an approach. A comparison with another method have also been
barely evoked by the authors, who mentioned however a significant
speed-up provided by Broyden algorithm for large $N_{\rm L}$ atomic
models. In particular, being contemporary with the publication of
Rybicki \& Hummer (1991), it is a pity that no comparison with the
MALI method could be made yet. Such an evaluation is the scope of the
present work.


\section{The numerical scheme}

For a $N_{L}$-level atomic model, the ESE will in general write as a
set of elementary equations:

\begin{eqnarray}
\sum\limits_{j<i} [n_i A_{ij}-(n_j B_{ji}-n_i B_{ij}) \bar{J}_{ij}]\nonumber\\
-\sum\limits_{j>i} [n_j A_{ji}-(n_i B_{ij}-n_j B_{ji})\bar{J}_{ij}]\nonumber\\
+\sum\limits_{j\neq i} (n_i C_{ij}-n_j C_{ji})=0
\label{eq:ese}
\end{eqnarray}
where the $A_{ij}$ and $B_{ij}$ stand respectively for the spontaneous
emission, and the absorption and stimulated emission rates, $n_i$
represents the population density for each energy level, and
$\bar{J}_{ij}$ is the scattering integral for each radiatively allowed
transition we shall consider. Besides the radiative processes, the
$C_{ij}$ are collisional excitation and de-excitation rates. In
general, these rates depend on the electronic density so that, if the
latter is not known a priori, terms like $n_i C_{ij}$ are nonlinear in
the population densities. Hereafter we shall consider only cases for
which the collisional rates are known a priori.

The scattering integral entering the ESE is formally written as:

\begin{equation}
\bar{J}_{ij} = \Lambda_{ij} [S_{ij}] \, ,
\end{equation}
where, assuming complete redistribution in frequency, the source
function is defined as:

\begin{equation}
{S}_{ij} = { {n_i A_{ij}} \over {n_j B_{ji}-n_i B_{ij}}} \, .
\end{equation}

A large system of Eqs. (\ref{eq:ese}) is homogeneous so,
practically, one of the equations have to be replaced by a constraint
equation like, for instance, a conservation equation of the form:

\begin{equation}
\sum\limits_{j=1}^{N_L}n_j = n_t \, .
\label{eq:cons}
\end{equation}

\begin{figure}
  \centering
  \includegraphics[width=9cm,angle=0]{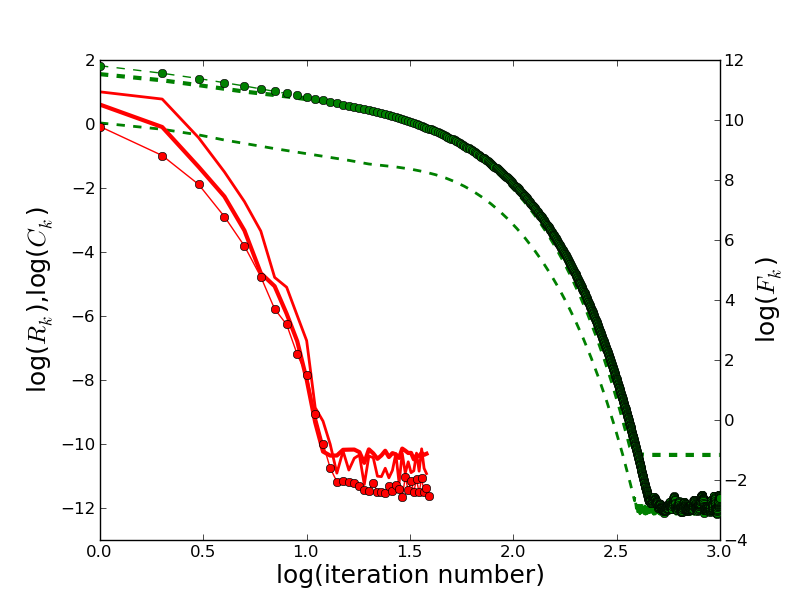}
  \caption{Typical relative error, $R_k$ (thin), convergence error,
    $C_k$ (thick) and $F_k$ (dotted) respectively for MALI (dash) and
    Broyden (full) vs. the number of iteration, for both scheme.}
  \label{Fig1}
\end{figure}

\subsection{Broyden's algorithm}

The system of Eqs. (\ref{eq:ese}) and (\ref{eq:cons}) can be
reformulated by defining a {\em function} $F$ acting on the set of
$n^{(\tau)}=(n_1^{(\tau)},...,n_i^{(\tau)},...,n_{N_L}^{(\tau)})$
where $\tau$ is a discrete depth along the opacity scale used to
sample our slab or atmosphere. $F$ is defined such as:

\begin{eqnarray}
\lefteqn \,F_i^{(\tau)}= & \sum\limits_{j<i} & [n_i^{(\tau)}A_{ij}-(n_j^{(\tau)}B_{ji}-n_i^{(\tau)}B_{ij})\bar{J}_{ij}]\nonumber\\
 & -\sum\limits_{j>i} & [n_j^{(\tau)}A_{ji}-(n_i^{(\tau)}B_{ij}-n_j^{(\tau)}B_{ji})\bar{J}_{ij}]\nonumber\\
 & +\sum\limits_{j\neq i} & (n_i^{(\tau)}C_{ij}-n_j^{(\tau)}C_{ji})
\label{eq:F1}
\end{eqnarray}
for $i \neq N_{\rm L}$  and, if $i=N_{\rm L}$

\begin{equation}
F_i^{(\tau)}=\sum\limits_{j=1}^{N_L}n_j^{(\tau)}-n_t \, .
\label{eq:F2}
\end{equation}
Computations of $F$ requires repeated evaluations of scattering
  integrals $\bar{J}_{ij}$ -- Eq. (2) -- that we perform here using
  the well-know short-characteristics method with monotonic parabolic
  interpolation introduced by Auer \& Paletou (1994). 

In that frame, and using the Sherman-Morrison formula (see e.g., Press
et al. 1992) which provides an analytical formula for the direct
computation of the {\em inverse} of the Broyden matrix, our algorithm
consists in the following steps. We choose an initial vector $n_0$, at
every $\tau$ depth, and an initial Broyden matrix $B_0$; then we
compute $B_{0}^{-1}$. The iterative scheme is such that:

\begin{equation}
\delta n_k=-B_k^{-1}F(n_k) \, ,
\end{equation}
then update

\begin{equation}
n_{(k+1)}=n_k+ \delta n_k \, ,
\end{equation}
then compute

\begin{equation}
\delta F_k=F(n_{(k+1)})-F(n_k) \, ,
\end{equation}
and finally, update

\begin{equation}
B_{(k+1)}^{-1}=B_k^{-1}+\frac{(\delta n_k-B_k^{-1} \delta F_k)\delta n_k^T B_k^{-1}}{(\delta n_k^TB_k^{-1})\delta F_k}
\end{equation}
while ${\| F \|_{2}} > \varepsilon$. Practically, we have chosen
$\varepsilon = 10^{-2}$, which guarantees that we have reached a fully
converged state (see also Fig.\,\ref{Fig1}).

\begin{figure}
  \centering
  \includegraphics[width=9cm,angle=0]{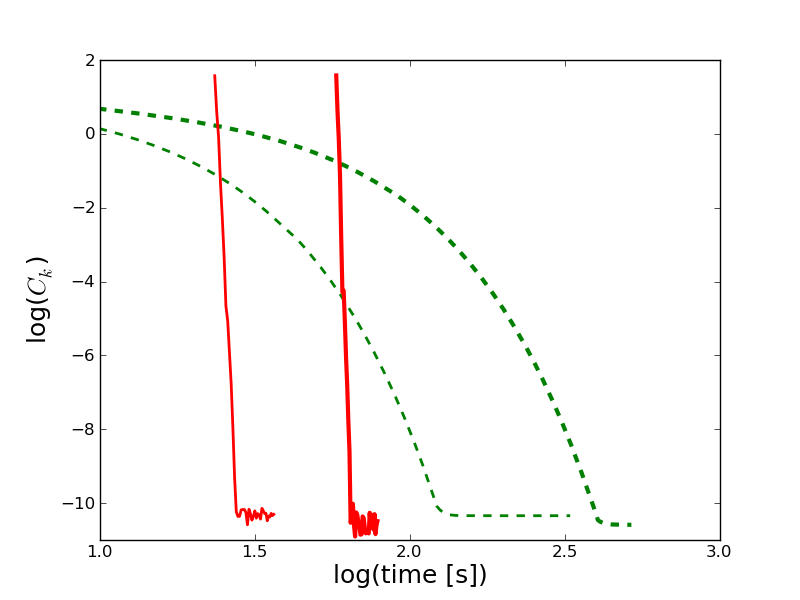}
  \caption{The convergence error, $C_k$, for MALI (dashed) and Broyden
    (full) are displayed vs. computing time for a 5-level H atom and,
    respectively, 5 (thin) and 8 (thick) points per decade in optical
    depth. }
  \label{Fig2}
\end{figure}

\subsection{Initialization}

The proper initialization of the Broyden scheme is a critical
issue. We employed the following method which was tested as suitable,
both from the standpoint of an adequate start and from the one of an
acceptable computing time.

Before starting the iterative scheme, we assume LTE populations for
our model-atom. In such a way, the grand function $F$ defined by
Eqs. (\ref{eq:F1}) and (\ref{eq:F2}) can be fully evaluated. Then, we
compute an initial Jacobian, $B_0$, using the finite difference scheme
{\tt fdjac} described by Press et al. (1992).

In the following figures relative to the timing properties of
Broyden's method vs. MALI, we shall always include the specific time
necessary for the evaluation of $B_0$.


\section{Comparison of Broyden vs. MALI}

\begin{figure}
  \centering
  \includegraphics[width=9cm,angle=0]{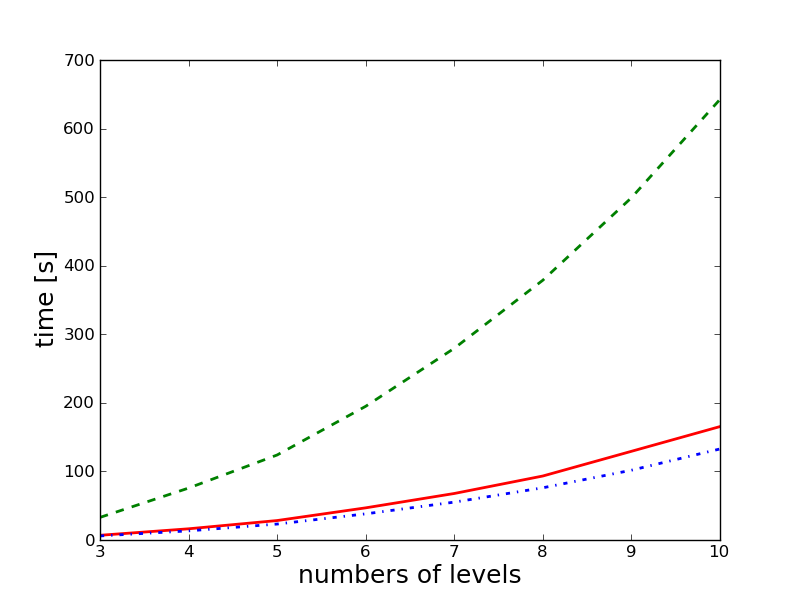}
  \caption{Respective computing times vs. number of levels of the
    H-atom model, for for MALI (dashed) and Broyden (full). The
    dash-dotted curve corresponds to the time required for the
    evaluation of the initial Jacobian.}
  \label{Fig3}
\end{figure}

 We adopted the popular ``flavor'' of the MALI method, using a
  diagonal approximate operator, as described by Rybicki \& Hummer
  (1991), and without acceleration of convergence schemes.  

In order to compare the properties of the Broyden scheme vs. MALI, we
adopted a 1D semi-infinite plane parallel slab model of $\tau_{\rm
  max}=10^{10}$, discretized in various number of points
per decade in optical depth, using also 3 to 10 energy levels H-atom,
inspired by the classic benchmark proposed by Avrett (1968; see
also L\'eger \& Paletou 2007). As in the latter, the slab temperature was
fixed at 5000\,K and the collisional rates set at $10^5 \,{\mathrm s}^{-1}$.
 We also adopted the definitions initially proposed by Auer et
al. (1994) for the relative error, from an iteration $(k)$ to another:

\begin{equation}
R_k = {\left| \left| { {n_{(k)} - n_{(k-1)}} \over {n_{(k)}} } \right| \right|_{\infty}} \, ,
\end{equation}
and for the ``convergence error'':

\begin{equation}
C_k = {\left| \left| { {n_{(k)} - n_{(\infty)}} \over {n_{(\infty)}} } \right| \right|_{\infty}} \, ,
\end{equation}
where $n_{(\infty)}$ is the ``fully converged'' solution obtained, for
a given method and model, after a large number of iterations. We also
introduce the quantity

\begin{equation}
F_k = {\| F_{(k)} \|_{2}}
\end{equation}
i.e., the Euclidian norm of $F$, the function defined by
Eqs. (\ref{eq:F1}) and (\ref{eq:F2}).
Note that $F_k^{(M)}$ for the MALI method is defined by a modified
Eq. (\ref{eq:F1}) following the preconditioning strategy proposed by
Rybicky \& Hummer (1991).


\subsection{Convergence}

In Fig.\,\ref{Fig1}, we display the rates of convergence of the
Broyden and MALI methods, respectively. The convergence error $C_k$
for each method have been computed using population densities obtained
once $F_k < {10}^{-2}$ in both cases. The case used here is a 5-level
H atom with a 1D slab discretized by a 5 points per decade grid in
optical depth. It is worth noting that, in order to reach $F_k <
{10}^{-2}$ and a well-converged solution though, one should iterate up
to reaching $R_k$ values as small as ${10}^{-10}$ typically.  In terms
of number of iterations, Broyden typically beats MALI of more than one
order of magnitude. However, the quite distinct nature of each method
makes such a comparison incomplete. Hereafter, we carry on such an
analysis but we shall compare respective {\em computing times}.

\subsection{Sensitivity to the spatial (optical depth) refinement}

In Fig.\,\ref{Fig2}, we turn to an analysis of the respective timing
properties of Broyden and MALI. It is shown that Broyden, again,
always beats MALI by a typical factor of the order of about 4-5 in
time. It is also the case when the spatial grid refinement is
increased from 5 to 8 points per decade, for instance. It is important
to note that timings given for Broyden {\em include} the computation
of the initial matrix $B_0$. This is why the rates of convergence
displayed for the Broyden method do not start at $t=0$ in
Fig.\,\ref{Fig2}.

\subsection{Sensitivity to the number of transitions}

A next important point to investigate relates to the advantage of
Broyden against MALI when an increasing number of atomic transitions
is considered.  Again, as demonstrated in Fig.\,\ref{Fig3}, the full
Broyden iterative process is always significantly faster than MALI. In
general, the gain due to the Broyden method is of the order of 4-5 in
total computing time. This is less than the the gain of the order of 8
already reported by Koesterke et al. (1992), although their method of
reference was presumably different from MALI.

\subsection{Discussion}

  We are aware that the MALI method can be speeded-up by acceleration
  of convergence schemes (see e.g., Auer 1991). However the most
  significant improvements in the field of iterative methods for the
  non-LTE radiative transfer problem were brought by the introduction
  of the Gauss-Seidel (GS) and Successive Over-Relaxation (SOR)
  methods (Trujillo Bueno \& Fabiani Bendicho 1995). It was already
  shown, for instance, that SOR always beats both Jacobi (i.e.,
  accelerated $\Lambda$-iteration with the diagonal of the full
  $\Lambda$ operator as an approximate operator) and GS, even when Ng
  acceleration of convergence is applied.

  Beyond the fact that Broyden is significantly faster than MALI, we
  can also add that Broyden is {\em as competitive} as the SOR method,
  according to Paletou \& L\'eger (2007; see their Table 1 where
  comparable timing and the corresponding iteration numbers were given
  for MALI, as we used it in the present study, GS and SOR).

  The Broyden method is also potentially more advantageous than MALI
  and GS/SOR, because of its intrinsic capability to deal with the
  self-consistent evaluation of the electron density in a multilevel
  non-LTE problem, if necessary -- a problem for which MALI needs to
  plug-in a Newton-Raphson scheme to it, as proposed by Heinzel (1995)
  and Paletou (1995).

  Another important point is that, as indicated in our Fig. 3, a great
  deal of time of our Broyden code is spent in the computation of the
  initial Jacobian, a task which can be performed with great advantage
  using parallel computing. The inner structure of the {\tt fdjac}
  routine permits, indeed, parallelization with a high scalability.

  As a final comment, it is also important to consider that Broyden's
  method can be easily implemented in already existing codes, {\em
    without the need of modifying the formal solver}, unlike with
  GS/SOR methods.

%
%
%


\section{Conclusion}

We propose an alternative method for the solution of the non--LTE
multilevel radiation transfer problem. It is based on Broyden's method
for the solution of nonlinear systems of equations.  The method is
easy to implement and it is about of factor of 4.5 times faster than
the well-known MALI method. Another advantage is that it does not
require any modification of usual formal solvers, as it is the case
for GS-SOR methods developed after MALI. It is also potentially very
well-suited for parallel computing. Further tests will include the
self-consistent treatment of the ionization balance, usually treated
together with MALI with a Newton-Raphson scheme. In a next step, we
shall consider more demanding models such as H$_{2}$O, for instance.

\end{document}